\documentstyle[12pt]{article}
\catcode`@=11
\newcount\@tempcntc
\def\@citex[#1]#2{\if@filesw\immediate\write\@auxout{\string\citation{#2}}\fi
  \@tempcnta\z@\@tempcntb\m@ne\def\@citea{}\@cite{\@for\@citeb:=#2\do
    {\@ifundefined
       {b@\@citeb}{\@citeo\@tempcntb\m@ne\@citea\def\@citea{,}{\bf ?}\@warning
       {Citation `\@citeb' on page \thepage \space undefined}}%
    {\setbox\z@\hbox{\global\@tempcntc0\csname b@\@citeb\endcsname\relax}%
     \ifnum\@tempcntc=\z@ \@citeo\@tempcntb\m@ne
       \@citea\def\@citea{,}\hbox{\csname b@\@citeb\endcsname}%
     \else
      \advance\@tempcntb\@ne
      \ifnum\@tempcntb=\@tempcntc
      \else\advance\@tempcntb\m@ne\@citeo
      \@tempcnta\@tempcntc\@tempcntb\@tempcntc\fi\fi}}\@citeo}{#1}}
\def\@citeo{\ifnum\@tempcnta>\@tempcntb\else\@citea\def\@citea{,}%
  \ifnum\@tempcnta=\@tempcntb\the\@tempcnta\else
   {\advance\@tempcnta\@ne\ifnum\@tempcnta=\@tempcntb \else \def\@citea{--}\fi
    \advance\@tempcnta\m@ne\the\@tempcnta\@citea\the\@tempcntb}\fi\fi}
\catcode`@=12
\def\theequation{\arabic{section}.\arabic{equation}}
\begin{document}

\begin{flushright}
RAL/95--024\\
MPI/PhT/95--22\\
ZTF/95--02\\
March 1995
\end{flushright}

\begin{center}
{\Large{\bf Semileptonic lepton-number/flavour-violating}}\\[0.3cm]
{\Large{\bf {$\tau$} decays in Majorana neutrino
models}}\\[1.5cm]
{\large A.~Ilakovac}$^a${\large ,~B.A.~Kniehl}$^b${\large ,
and~A.~Pilaftsis}$^c$\\[0.4cm]
{\em $^a$Faculty of Science, Department of Physics, Bijeni\v cka 32,
41001 Zagreb, Croatia}\\[0.2cm]
{\em $^b$Max-Planck-Institut f\"ur Physik, F\"ohringer Ring 6,
80805 Munich, Germany}\\[0.2cm]
{\em $^c$Rutherford Appleton Laboratory, Chilton, Didcot, Oxon, OX11 0QX,
UK}
\end{center}
\vskip1.5cm
\centerline{\bf ABSTRACT}
Motivated by the recent investigation of neutrinoless $\tau$-lepton
decays by the CLEO collaboration, we perform a systematic analysis
of such decays in a possible new-physics scenario with heavy Dirac/Majorana
neutrinos, including heavy-neutrino nondecoupling effects, finite quark
masses, and quark as well as meson mixings.
We find that $\tau$-lepton decays into an electron or muon
and a pseudoscalar or vector meson can have branching ratios
close to the experimental sensitivity.
Numerical estimates show that the predominant decay modes of this kind are
$\tau^-\to e^-\phi$, $\tau^-\to e^-\rho^0$, and $\tau^-\to e^-\pi^0$,
with branching ratios of order $10^{-6}$.

\newpage
\section{Introduction}
\setcounter{equation}{0}
\indent

Recently, the CLEO collaboration has reported their experimental results on
22 neutrinoless decay channels of the $\tau$ lepton that violate
lepton flavour and/or lepton number \cite{CLEO}.
Candidates for lepton-flavour/number-violating events have been found in
the decays $\tau^-\to e^-\bar{K}^{*0}$, $\tau^-\to e^-\pi^+K^-$,
$\tau^-\to \mu^-\pi^-K^+$, and $\tau^-\to \mu^+\pi^-K^-$. Such decays
are strictly forbidden in the minimal Standard Model (SM) due to the fact
that the light neutrinos, $\nu_e$, $\nu_\mu$, and $\nu_\tau$, are exactly
massless, so that chirality conservation implies lepton number/flavour
conservation to all orders of the perturbative expansion.
Since there is no fundamental theoretical reason for lepton number/flavour
conservation in nature, future confirmation of the CLEO candidates
may point towards physics beyond the SM and, in particular, to some
modification of the lepton sector.
Such possible lepton-number/flavour-violating effects can naturally be
accounted for in the context of leptoquark models~\cite{PS},
left-right-symmetric models~\cite{MS}, $R$-parity-violating supersymmetric
scenarios~\cite{Rparity}, or theories containing heavy Dirac and/or Majorana
neutrinos~\cite{SV,ZPC}.

In this paper, we will study the size of new-physics interactions
in models with heavy Dirac and/or Majorana neutrinos. In such scenarios,
decays of the $\tau$ lepton into three charged leptons, such as
$\tau\to eee$, {\it etc.}, have been analyzed in Ref.~\cite{IP}. Here,
our main interest will be devoted to semileptonic decays of the $\tau$ lepton.
Specifically, we will analyze decays of the type $\tau^-\to e^-\pi^0$,
$\tau^-\to e^-\rho^0$, {\it etc.} In a previous work~\cite{GGV},
three of the numerous decay channels of this type were considered in the
framework of a theory with heavy Dirac neutrinos. We will extend that analysis
by including lepton-number-violating interactions due to Majorana neutrinos,
heavy-neutrino nondecoupling effects~\cite{IP,AP},
finite-quark-mass contributions, Cabibbo-Kobayashi-Maskawa (CKM) quark
mixings, and meson mixings.
We will perform a complete analysis, which
comprises all the ten decay channels of the type
$\tau^-\to e^-M^0$,
where $M^0$ denotes either a pseudoscalar or vector meson.
The effect of a modified lepton sector on $\tau$ decays into
two mesons, {\em i.e.}, $\tau^-\to l^\mp M_1^- M_2^\pm$, will be
estimated in a separate communication \cite{amon}.

In our calculations, we will adopt the conventions and the model
described in Ref.~\cite{ZPC}. In this minimal model, which is based on the SM
gauge group, the neutrino sector is extended by the presence of a number
($n_R$) of neutral isosinglets leading to $n_R$ heavy Majorana neutrinos
($N_j$), while the quark sector retains the SM structure. If the theory
contains more than one neutral isosinglet, then the heavy-light neutrino
mixing~\cite{LL},
\begin{equation}
\label{constr}
(s^{\nu_l}_L)^2\ \equiv\ 1- \sum\limits_{i=1}^3 |B_{l\nu_i}|^2\
=\ \sum\limits_{j=1}^{n_R}|B_{lN_j}|^2,
\end{equation}
scales as $(m^\dagger_D(m_M^{-1})^2 m_D)_{ll}$, where $m_D$ is the
Dirac mass matrix related to the breaking of the SU(2)$_L$
gauge symmetry and $m_M$ is a general $n_R\times n_R$
isosinglet mass matrix.
Light neutrinos ($\nu_\l$) can radiatively acquire masses in compliance
with experimental upper bounds~\cite{ZPC}, whereas the
lepton-flavour-violating mixings $(s^{\nu_l}_L)^2$ are dramatically
relaxed and do not obey the traditional seesaw suppression relation,
$(s^{\nu_l}_L)^2\propto m_\nu/m_N$. Then, the
$(s^{\nu_l}_L)^2$ may be viewed as free phenomenological parameters,
which may be constrained by a variety of low-energy data~\cite{LL,BGKLM}.
Throughout this paper, we will consider the following conservative upper
limits for the flavour-violating mixings~\cite{BGKLM}:
\begin{equation}
\label{upli}
(s^{\nu_e}_L)^2, (s^{\nu_\mu}_L)^2 < 0.015,\qquad (s^{\nu_\tau}_L)^2<0.050,
\qquad (s^{\nu_e}_L)^2(s^{\nu_\mu}_L)^2<10^{-8}.
\end{equation}
Note that these upper bounds are sensitive, to a large extent, to
the degree of confidence level (CL) considered in the global analyses
({\em i.e.}, 95\% or 99\% CL) and further model-dependent
assumptions~\cite{IP}.

In an $n_G$-generation model, the couplings of the charged- and
neutral-current interactions are correspondingly mediated by the mixing
matrices
\begin{equation}
B_{lj}\ =\ \sum\limits_{k=1}^{n_G}V^l_{lk}U^{\nu\ast}_{kj},\qquad
C_{ij}\ =\ \sum\limits_{k=1}^{n_G}U^\nu_{ki}U^{\nu\ast}_{kj},
\end{equation}
where $V^l$ and $U^\nu$ are the unitary matrices that are needed to
diagonalize the charged-lepton and neutrino mass matrices, respectively.
$B$ and $C$ satisfy a number of identities that
guarantee the renormalizability of the model~\cite{ZPC,KP}.
Such identities are found to be very helpful in order to reduce
the number of free parameters present in such theories and, by the same token,
to establish relations between $B$, $C$, and the heavy-neutrino masses.
For definiteness, in our numerical calculations, we will use
a model with two right-handed neutrinos.
In such a scenario, we have~\cite{IP}
%\begin{equation}
%B_{lN_1}\ =\ \frac{\rho^{1/4} s^{\nu_l}_L}{\sqrt{1+\rho^{1/2}}}\ , \qquad
%B_{lN_2}\ =\ \frac{i s^{\nu_l}_L}{\sqrt{1+\rho^{1/2}}}\ , \label{BlN}
%\end{equation}
\begin{equation}
B_{lN_1}\ =\ \frac{\rho^{1/4} s^{\nu_l}_L}{\sqrt{1+\rho^{1/2}}}\ , \qquad
B_{lN_2}\ =\ i\rho^{-1/4}B_{lN_1}, \label{BlN}
\end{equation}
where $\rho=m^2_{N_2}/m^2_{N_1}$, with $N_1$ and $N_2$ being the
heavy Majorana neutrinos.
Furthermore, the mixings $C_{N_iN_j}$ are given by
\begin{eqnarray}
C_{N_1N_1} &=& \frac{\rho^{1/2}}{1+\rho^{1/2}}\ \sum\limits_{i=1}^{n_G}
(s^{\nu_i}_L)^2, \qquad C_{N_2N_2}\ =\ \rho^{-1/2}C_{N_1N_1},\nonumber\\
C_{N_1N_2}&=& -C_{N_2N_1}\ =\ i\rho^{-1/4}C_{N_1N_1}.
\end{eqnarray}
Obviously, our minimal scenario only depends on
$m_{N_1}$ and $m_{N_2}$---or, equivalently, on
$m_{N_1}$ and $\rho$---, and $(s^{\nu_l}_L)^2$,
which are assumed to satisfy the constraints in Eq.~(\ref{upli}).

The outline of this work is as follows. In Section~2, we will calculate
analytically the branching ratios of the decay processes $\tau^-\to
e^-M^0$.
Technical details will be relegated to the Appendix.
Our numerical results will be presented in Section~3.
Section~4 contains our conclusions.

\section{{\boldmath $\tau^-\rightarrow l'^-M^0$}}
\setcounter{equation}{0}

Charge conservation forbids the lepton-number-violating decays of a
$\tau$ lepton into a meson and an antilepton. For the same reason,
the outgoing meson has to be neutral.
The recent CLEO experiment \cite{CLEO} observes an event for the
decay $\tau^-\rightarrow e^-\bar{K}^{0*}$ within the signal region,
which is still consistent with the estimated background due to
hadron misidentification.
The same experiment has considerably lowered the upper bounds on the rates
of the decays with one $\rho^{0}$ or one $K^{0*}$ in the final state.

The scattering-matrix element of $\tau^-\to l'^- M^0$ receives contributions
from $\gamma$-exchange graphs, $Z$-boson-exchange graphs, and box graphs,
\begin{equation}
S(\tau^-\to l'^- M^0)=S_\gamma(\tau^-\to l'^- M^0)+S_Z(\tau^-\to l'^- M^0)
+S_{Box}(\tau^-\to l'^- M^0)\, .
\end{equation}
Feynman diagrams pertinent to these decays are shown in Fig.~1.
The $\gamma$ and $Z$-boson amplitudes factorize into leptonic vertex
corrections and hadronic pieces.
The loop integrations are straightforward.
The hadronic parts are local.
Exploiting translation invariance, the phases that describe the
centre-of-mass motion of $M^0$ may be isolated and one is left with
space-time independent hadronic matrix elements.
These phases assure four-momentum conservation. The $\gamma$ and $Z$-boson
amplitudes read
\begin{eqnarray}
S_\gamma(\tau^-\to l'^- M^0) &=&
(2\pi)^4\,\delta^{(4)}(p-p'-p_M)\,
\frac{i\alpha^2_W s_W^2}{4M_W^2}\,
\bar{u}_{l'}\Big[F_\gamma^{\tau l'}(\gamma^\mu-
   \frac{q^\mu\not\!q}{q^2})(1-\gamma_5)
\nonumber\\
&&-G_\gamma^{\tau l'}\frac{i\sigma^{\mu\nu}q_\nu}{q^2}
   (m_\tau (1+\gamma_5)+m_{l'}(1-\gamma_5))\Big]u_\tau
\nonumber\\
&&\times \langle M^0|\frac{2}{3}\bar{u}(0)\gamma_\mu u(0)
           -\frac{1}{3}\bar{d}(0)\gamma_\mu d(0)
-\frac{1}{3}\bar{s}(0)\gamma_\mu s(0) |0\rangle,\qquad
\qquad
\nonumber\\
S_Z(\tau^-\to l'^- M^0) &=&
(2\pi)^4\,\delta^{(4)}(p-p'-p_M)\:
\frac{i\alpha^2_W}{16M_W^2}\, F_Z^{\tau l'}\,
\bar{u}_{l'}\gamma^\mu(1-\gamma_5)u_\tau
\nonumber\\
&&\times  \Bigg( \langle M^0|\bar{u}(0) \gamma_\mu \Big(1-\gamma_5
-\frac{8}{3}s_W^2\Big) u(0)|0\rangle\nonumber\\
&&-\langle M^0|\bar{d}(0)\gamma_\mu\Big( 1-\gamma_5
-\frac{4}{3}s_W^2\Big)d(0)|0\rangle \nonumber\\
&&-\langle M^0 | \bar{s}(0)\gamma_\mu\Big( 1-\gamma_5
-\frac{4}{3}s_W^2\Big)s(0) |0\rangle\Bigg), \label{SZ}
\end{eqnarray}
where $p$, $p'$, and $p_M$ are the four-momenta of $\tau$, $l'$, and
$M^0$, respectively, $q=p-p'=p_M$,
$\alpha_W=\alpha_{\rm em}/\sin^2\theta_W\approx0.0323$ is the weak
fine-structure constant, and $u(x)$, $d(x)$, and $s(x)$ are
quark-field operators acting on the meson states $|M^0\rangle$.
In Eq.~(\ref{SZ}), $F_Z^{\tau l'}$, $F_\gamma^{\tau l'}$, and
$G_\gamma^{\tau l'}$ are form factors, which may be found in
Ref.~\cite{IP}.

The box diagram is more involved, as it contains a bilocal quark operator.
Taking the difference, $X$, and the averaged sum of the space-time coordinates
of the two hadronic vertices as integration variables, using
translation invariance, and performing the integration over the
leptonic space-time coordinates, one arrives at the following
expression for the box amplitude:
\begin{eqnarray}
S_{Box}(\tau^-\to l'^- M^0)&=&
(2\pi)^4\,\delta^{(4)}(p-p'-p_M)\:\frac{\alpha^2_W\pi^2}{2}
\sum_{i=1}^{n_G+n_R}B_{l'i}B^*_{\tau i}
\int \frac{d^4l}{(2\pi)^4}
\nonumber\\
&&
\times \bar{u}_{l'}\gamma^\mu\frac{\not\!p'+\not\!l}{(p'+l)^2-m_{n_i}^2}
  \gamma^\nu(1-\gamma_5)u_\tau
\ \frac{1}{l^2-M_W^2}\
  \frac{1}{(q-l)^2-M_W^2}
\nonumber\\
&&
\times \Big[\sum_{d_{a,b}=d,s}\sum_{i=1}^{n_G}V^*_{u_id_b}V_{u_id_a}
\int d^4X\: e^{-i(l-q/2)X}
\nonumber\\
&&\times
\langle M^0|\bar{d}_b\Big(\frac{X}{2}\Big)\gamma_\mu(1-\gamma_5)S_F^{(u_i)}(X)
                         \gamma_\nu(1-\gamma_5)d_a
\Big(-\frac{X}{2}\Big)|0\rangle
\nonumber\\
&&-
\sum_{i=1}^{n_G} V^*_{ud_i}V_{ud_i}
\int d^4X\: e^{-i(l-q/2)X}
\nonumber\\
&&
\times \langle M^0|\bar{u}\Big(-\frac{X}{2}\Big)\gamma_\nu(1-\gamma_5)
                  S_F^{(d_i)} (-X)
                         \gamma_\mu(1-\gamma_5)u
\Big(\frac{X}{2}\Big)|0\rangle
\Big]\nonumber\\
&&+\quad (W\rightarrow G), \label{SSbox}
\end{eqnarray}
where $(W\rightarrow G)$ stands for the terms obtained
by replacing one or two $W$ bosons with unphysical charged Higgs bosons and
$S_F^{(u_i,d_i)} (x) $
are the $u_i$- and $d_i$-quark propagators in coordinate space.
An exact expression for the quark propagator is not known,
but using the free-quark propagator is known to yield a good approximation
for large momentum transfers.
The corresponding box amplitude which contains only free quarks, receives
its dominant support from momenta with virtualities comparable to $M_W$
(if the virtual lepton is light) or larger.
Only a small fraction of the
Feynman-parameter space permits potentially large contributions from low
squared-momentum values. Therefore, it is plausible to approximate
the quark propagator with the free-particle propagator and
neglect the momenta of the external leptons
and the $X$ dependence of the quark wave functions as well.
Thus, we recover the free-quark expressions for the box functions and
evaluate the hadronic matrix elements by taking the quark current operators
to be local. In this way, Eq.~(\ref{SSbox}) simplifies to
\begin{eqnarray}
S_{Box}(\tau^-\to l'^- M^0)&=&
(2\pi)^4\,\delta^{(4)}(p -p'-p_M)\:
\frac{i\alpha^2_W}{16M_W^2}\,
\bar{u}_{l'}\gamma_\mu (1-\gamma_5)u_\tau
\nonumber\\
&&
\times \Big[F_{Box}^{\tau l'uu}
    \langle M^0|\bar{u}(0)\gamma^\mu (1-\gamma_5)u(0)|0\rangle
\nonumber\\
&&-\sum_{d_{a,b}=d,s}F_{Box}^{\tau l'd_ad_b}
    \langle M^0|\bar{d}_a(0)\gamma^\mu (1-\gamma_5)d_b(0)|0\rangle\Big],
\end{eqnarray}
where $F_{Box}^{\tau l'd_ad_b}$ and $F_{Box}^{\tau l'uu}$ may
be found in Appendix~A.

To calculate hadronic matrix elements, we invoke the hypothesis of the
partial conservation of axial vector currents (PCAC) \cite{PCAC,VMD,EVMD},
\begin{equation}
A^P_\mu(x) = i\sqrt{2}f_P\partial_\mu P(x)+\cdots,\label{PCAC}
\end{equation}
where the dots denote terms not contributing to the meson--vacuum amplitude,
$f_P$ is the decay constant of the pseudoscalar meson $P$,
represented by the field $P(x)$, and $A^P_\mu(x)$ is the axial-vector current
having the same quark content as $P$.
The pion decay constant is $f_\pi=92$~MeV.
Furthermore, we exploit  the vector-meson dominance (VMD)
relation \cite{VMD,EVMD},
\begin{equation}
\label{VMD}
j_{em}^\mu(x)=
\frac{m_\rho^2}{2\gamma_\rho}\rho^\mu (x)
+\frac{m_\omega^2}{2\sqrt{3}\gamma_\omega}\omega^\mu (x) \sin\theta_V
+\frac{m_\phi^2}{2\sqrt{3}\gamma_\phi}\phi^\mu (x) \cos\theta_V,
\end{equation}
and its extension for any vector current \cite{EVMD},
\begin{equation}
\label{EVMD}
V_\mu^{\tilde{V}}(x)=\frac{m_{\tilde{V}}^2}{\sqrt{2}\gamma_{\tilde{V}}}
\tilde{V}_\mu(x).
\end{equation}
In Eq.~(\ref{VMD}), $j_{em}^\mu(x)$ is the electromagnetic current,
$\rho^\mu(x)$, $\omega^\mu(x)$, and $\phi^\mu(x)$ are
the $\rho$-, $\omega$-, and $\phi$-meson fields, respectively,
$\gamma_\rho$, $\gamma_\omega$, and $\gamma_\phi$ measure the strengths
of their couplings to the photon, and $\theta_V$ is the usual mixing angle
of the octet and singlet vector-meson states. In Eq.~(\ref{EVMD}),
$V_\mu^{\tilde{V}}$ is the vector field having the same quark content
as the vector meson field $\tilde{V}$.
Equation~(\ref{EVMD}) is based on the assumption that the dominant
contribution to the form factors is due to the vector mesons,
which works very well for the electromagnetic current \cite{KRS_VMD}.

The calculation of the hadronic matrix elements proceeds as follows.
One expresses the quark operators, which appear in the hadronic
matrix elements, in terms of the axial-vector [$A^P_\mu(x)$] and vector
currents [$V^{\tilde{V}}_\mu(x)$] that have the same quark content as the
produced pseudoscalar ($P$) and vector mesons ($\tilde{V}$).
Then, one applies Eqs.~(\ref{PCAC}) and (\ref{EVMD}).
The relevant matrix elements read
\begin{eqnarray}
\label{matels}
\langle 0|A^P_\mu(x)|M(p_M)\rangle&=&
\delta_{MP}\sqrt{2} f_P p_{P\mu} e^{-ip_Px},\nonumber\\
\langle 0|V^{\tilde{V}}_\mu(x)|M(p_M)\rangle&=&
\delta_{M\tilde{V}}
\frac{m_{\tilde{V}}^2}{\sqrt{2}\gamma_{\tilde{V}}}
  \epsilon_{\tilde{V}\mu}(p_{\tilde{V}}, \lambda_{\tilde{V}})
e^{-ip_{\tilde{V}}x},
\end{eqnarray}
where $\epsilon_{\tilde{V}\mu}$ stands for the polarization vector of the
vector boson $\tilde{V}$, and the Kronecker symbols, $\delta_{MP}$ and
$\delta_{M\tilde{V}}$, assure that the matrix elements give non-zero
contributions only if the final-state quantum numbers match those of the
vector and axial-vector currents.
The matrix elements appropriate to mesons in the final state emerge from
Eq.~(\ref{matels}) by Hermitean conjugation.

The decomposition of the vector and axial-vector
currents into meson field operators
depends on the quark content of the meson (for the pseudoscalar mesons,
see Table~I).
The quark content of pseudoscalar mesons having zero isospin and zero
hypercharge is not yet definitely established \cite{PDG,ZielinskiRosner}.
The mixing of SU(3)-octet and SU(3)-singlet meson states with zero isospin
and zero hypercharge is usually parameterized by some angle, $\theta_P$,
which is not precisely known.
The corresponding mixing angle for vector mesons is called $\theta_V$.
{}From the study of $\phi$ decays it is known that $\theta_V$ is very close to
the ideal value $\arctan(1/\sqrt{2})$.
Notice that the state $|M\rangle$ and the corresponding field
$M(x)$ have opposite quantum numbers. This is due to the convention
$\langle 0|M(x)|M(p)\rangle =e^{-ipx}\epsilon_M(p)$,
{\it i.e.}, the meson field annihilates the corresponding meson state.
In the second line of Table~I, we indicate the relevant creation and
annihilation operators that are contained in the meson states.
Here, $b_u$ and $d_s$ are the annihilation operators of the quark $u$ and
the antiquark $\bar s$,
respectively, and $b_u^\dagger$ and $d_s^\dagger$ are their creation
operators.
The quark structure of the vector-meson states and fields may be read off
from Table~I after the replacements $K^\pm\to K^{\pm 0}$,
$\pi^\pm\to \rho^\pm$, $\pi^0\to \rho^0$, $K^0\to K^{0*}$,
$\bar{K}^0\to \bar{K}^{0*}$, $\eta_{8,1}\to \phi_{8,1}$,
$\eta\to \phi$, $\eta'\to \omega$, and $\theta_P\to \theta_V$.

Following the procedure outlined above, we obtain the following expressions
for the $\tau^-\rightarrow  l'^-M^0$ matrix elements:
\begin{eqnarray}
T(\tau^-\to l'^- K^0) &=& \frac{i\alpha^2_W}{16M_W^2}\,
       \bar{u}_{l'}\gamma_\mu(1-\gamma_5)u_\tau
       \; \sqrt{2}f_K p_K^\mu\; F^{\tau l'sd}_{Box},\nonumber\\
T(\tau^-\to l'^- \bar{K}^0) &=& -\frac{i\alpha^2_W}{16M_W^2}\,
       \bar{u}_{l'}\gamma_\mu(1-\gamma_5)u_\tau
       \; \sqrt{2}f_K p_K^\mu\; F^{\tau l'ds}_{Box},\nonumber\\
T(\tau^-\to l'^- \pi^0)&=&-\frac{i\alpha^2_W}{16M_W^2}\,
       \bar{u}_{l'}\gamma_\mu(1-\gamma_5)u_\tau
       \; f_\pi p_\pi^\mu \;
       [2F_Z^{\tau l'}+F^{\tau l'uu}_{Box}+F^{\tau l'dd}_{Box}],\nonumber\\
T(\tau^-\to l'^- \eta)&=&\frac{i\alpha^2_W}{16M_W^2}\,
       \bar{u}_{l'}\gamma_\mu(1-\gamma_5)u_\tau
       \; f_\eta p_\eta^\mu \;
       \Big[-(\frac{2c_P}{\sqrt{3}}+\frac{\sqrt{2}s_P}{\sqrt{3}})F_Z^{\tau l'}
       \nonumber\\ &&
      +(-\frac{c_P}{\sqrt{3}}+\frac{\sqrt{2}s_P}{\sqrt{3}})F^{\tau l'uu}_{Box}
      +(\frac{c_P}{\sqrt{3}}-\frac{\sqrt{2}s_P}{\sqrt{3}})F^{\tau l'dd}_{Box}
        \qquad
        \nonumber\\ &&
         -(\frac{2c_P}{\sqrt{3}}+\frac{\sqrt{2}s_P}{\sqrt{3}})F^{\tau
l'ss}_{Box}
       \Big],\nonumber\\
T(\tau^-\to l'^- \eta')&=&\frac{i\alpha^2_W}{16M_W^2}\,
       \bar{u}_{l'}\gamma_\mu(1-\gamma_5)u_\tau
       \; f_{\eta'} p_{\eta'}^\mu \;
       \Big[(\frac{\sqrt{2}c_P}{\sqrt{3}}-\frac{2s_P}{\sqrt{3}})F_Z^{\tau l'}
       \nonumber\\ &&
       -(\frac{\sqrt{2}c_P}{\sqrt{3}}+\frac{s_P}{\sqrt{3}})F^{\tau l'uu}_{Box}
       +(\frac{\sqrt{2}c_P}{\sqrt{3}}+\frac{s_P}{\sqrt{3}})F^{\tau l'dd}_{Box}
        \qquad
        \nonumber\\ &&
       +(\frac{\sqrt{2}c_P}{\sqrt{3}}-\frac{2s_P}{\sqrt{3}})F^{\tau l'ss}_{Box}
       \Big],\nonumber\\
T(\tau^-\to l'^- K^{0*})&=&-\frac{i\alpha^2_W}{16M_W^2}\,
       \bar{u}_{l'}\gamma_\mu(1-\gamma_5)u_\tau
       \;
       \frac{m^2_{K^{0*}}}{\sqrt{2}\gamma_{K^{0*}}}\epsilon^\mu_{K^{0*}}
       \; F^{\tau l'sd}_{Box},\nonumber\\
T(\tau^-\to l'^- \bar{K}^{0*})&=&\frac{i\alpha^2_W}{16M_W^2}\,
       \bar{u}_{l'}\gamma_\mu(1-\gamma_5)u_\tau
       \;
       \frac{m^2_{K^{0*}}}{\sqrt{2}\gamma_{K^{0*}}}\epsilon^\mu_{K^{0*}}
       \; F^{\tau l'ds}_{Box},\nonumber\\
T(\tau^-\to l'^- \rho^0)&=&\frac{i\alpha^2_W}{16M_W^2}\,
       \frac{m^2_{\rho_0}}{\gamma_{\rho_0}}
            \epsilon^{\mu}_{\rho^0}
       \;
       \Big\{2s^2_W
           \bar{u}_{l'}
  \Big[\, F_\gamma^{\tau l'}\,
(\gamma_\mu-\frac{q_\mu\not\!q}{q^2})(1-\gamma_5)
       \nonumber\\
   &&        -G_\gamma^{\tau l'}\, \frac{i\sigma_{\mu\nu}q^\nu}{q^2}
              \Big(m_\tau (1+\gamma_5)+m_{l'}(1-\gamma_5)\Big)
                \Big] u_\tau
       \nonumber\\
   && +
           \bar{u}_{l'}\gamma_\mu(1-\gamma_5)u_\tau \:
            [c_{2W} F_Z^{\tau l'}+\frac{1}{2}F^{\tau l'uu}_{Box}
             +\frac{1}{2}F^{\tau l'dd}_{Box}]\},\nonumber\\
T(\tau^-\to l'^- \phi)&=&\frac{i\alpha^2_W}{16M_W^2}\,
       \frac{m_\phi^2}{\gamma_\phi}\epsilon^\mu_{\phi}
       \;
       \Big\{\frac{2s^2_Wc_V}{\sqrt{3}}
            \: \bar{u}_{l'}
             \Big[\, F_\gamma^{\tau l'}\, (\gamma_\mu-\frac{q_\mu\not\!q}{q^2})
                  (1-\gamma_5)
       \nonumber\\
   &&        -G_\gamma^{\tau l'}\, \frac{i\sigma_{\mu\nu}q^\nu}{q^2}
              \Big(m_\tau (1+\gamma_5)+m_{l'}(1-\gamma_5)\Big)
                  \Big] u_\tau
       \nonumber\\
   && +
            \bar{u}_{l'}\gamma_\mu(1-\gamma_5)u_\tau\:
             \Big[(\frac{c_V}{\sqrt{3}}c_{2W} +\frac{s_V}{\sqrt{6}})F_Z^{\tau
l'}
              +(\frac{c_V}{2\sqrt{3}}-\frac{s_V}{\sqrt{6}})F^{\tau l'uu}_{Box}
       \nonumber\\
   &&         -(\frac{c_V}{2\sqrt{3}}-\frac{s_V}{\sqrt{6}})F^{\tau l'dd}_{Box}
              +(\frac{c_V}{\sqrt{3}}+\frac{s_V}{\sqrt{6}})F^{\tau l'ss}_{Box}
             \Big]\Big\},\nonumber\\
T(\tau^-\to l'^- \omega)&=&\frac{i\alpha^2_W}{16M_W^2}\,
        \frac{m_\omega^2}{\gamma_\omega}
                 \epsilon^\mu_{\omega}
        \;
        \Big\{\frac{2s^2_Ws_V}{\sqrt{3}}
            \:\bar{u}_{l'}
             \Big[\, F_\gamma^{\tau l'}\, (\gamma_\mu-\frac{q_\mu\not\!q}{q^2})
                       (1-\gamma_5)
       \nonumber\\
    &&       -G_\gamma^{\tau l'}\, \frac{i\sigma_{\mu\nu}q^\nu}{q^2}
              \Big(m_\tau (1+\gamma_5)+m_{l'}(1-\gamma_5)\Big)\:
                    \Big] u_\tau
       \nonumber\\
    && +
            \bar{u}_{l'}\gamma_\mu(1-\gamma_5)u_\tau\:
             \Big[(\frac{s_V}{\sqrt{3}}c_{2W}-\frac{c_V}{\sqrt{6}})F_Z^{\tau
l'}
              +(\frac{s_V}{2\sqrt{3}}+\frac{c_V}{\sqrt{6}})F^{\tau l'uu}_{Box}
       \nonumber\\
    &&        -(\frac{s_V}{2\sqrt{3}}+\frac{c_V}{\sqrt{6}})F^{\tau l'dd}_{Box}
              +(\frac{s_V}{\sqrt{3}}-\frac{c_V}{\sqrt{6}})F^{\tau l'ss}_{Box}
             \Big]\Big\},
\end{eqnarray}
where we have introduced the short-hand notations
$s_W=\sin\theta_W$, $c_W=\cos\theta_W$, $c_{2W}=\cos2\theta_W$, and similarly
for $\theta_P$ and $\theta_V$.

The branching ratios for pseudoscalar mesons can be compactly written
in the form
\begin{eqnarray}
\label{brp}
B(\tau^-\rightarrow l'^-M^0)
&=&\frac{1}{8\pi}\frac{m_\tau}{\Gamma_\tau}
\frac{\lambda^{\frac{1}{2}}(m_\tau^2,m_{l'}^2,m^2_{M^0})}{m_\tau^2}|a_{M^0}|^2
\frac{(m_\tau^2-m_{l'}^2)^2-m_{M^0}^2(m_\tau^2+m_{l'}^2)}{m_\tau^2},\qquad
\end{eqnarray}
where the form factors $a_{M^0}$ are listed in Appendix A,
$\Gamma_\tau=2.16\ 10^{-12}$ GeV is the total width of the $\tau$ lepton
measured experimentally, and $\lambda (x,y,z) = (x-y-z)^2-4yz$.
Similarly, in the case of the vector mesons, one finds
\begin{eqnarray}
\label{brv}
B(\tau^-\rightarrow l'^-M^0)
&=&\frac{1}{8\pi}\frac{m_\tau}{\Gamma_\tau}
\frac{\lambda^{\frac{1}{2}}(m_\tau^2,m_{l'}^2,m^2_{M^0})}{m_\tau^2}
\Big[|c_{M^0}|^2\Big\{-\frac{12 m_{l'}^2}{m_{M^0}^2}
\nonumber\\ &&
+\frac{(m_\tau^2+m_{l'}^2)\Big(2(m_\tau^2-m_{l'}^2)^2
       -m_{M^0}^2(m_\tau^2+m_{l'}^2)-m_{M^0}^4\Big)}{m_\tau^2 m_{M^0}^4}\Big\}
\nonumber\\ &&
+|a_{M^0}+b_{M^0}|^2
\frac{(m_\tau^2-m_{l'}^2)^2+m_{M^0}^2(m_\tau^2+m_{l'}^2)-2m_{M^0}^4}
     {m_\tau^2 m_{M^0}^2}
\nonumber\\ &&
-6 \Re e\{(a_{M^0}+b_{M^0})c_{M^0}^*\}\:
\frac{(m_\tau^2-m_{l'}^2)^2-m_{M^0}^2(m_\tau^2+m_{l'}^2)}
     {m_\tau^2 m_{M^0}^2}\Big],\qquad
\end{eqnarray}
where $a_{M^0}$, $b_{M^0}$, and $c_{M^0}$ may also be found in Appendix A.

\section{Numerical results}
\setcounter{equation}{0}

In our numerical analysis, we will assume that the SM is extended
by two right-handed neutrinos, as described in the Introduction.
The additional parameters in this scenario are the two heavy-neutrino masses,
$m_{N_1}$ and  $m_{N_2}$, and the three mixing angles,
$s_L^{\nu_e}$, $s_L^{\nu_\mu}$, and $s_L^{\nu_\tau}$.
These are free parameters of the model, which may be limited by experiment.
The upper bounds on $s_L^{\nu_e}$, $s_L^{\nu_\mu}$, and $s_L^{\nu_\tau}$
are given in Eq.~(\ref{upli}).
On the other hand, the perturbative unitarity relations,
\begin{equation}
\frac{\Gamma_{N_i}}{m_{N_i}}<\frac{1}{2},
\end{equation}
lead to a global upper bound on $m_{N_1}$,
\begin{equation}
m^2_{N_1}\leq\frac{2M_W^2}{\alpha_W}
           \frac{1+\rho^{-\frac{1}{2}}}{\rho^{\frac{1}{2}}}
           \Big[ \sum_i (s_L^{\nu_i})^2\Big]^{-1},\label{mnbound}
\end{equation}
where $\rho$ is defined after Eq.~(\ref{BlN}) and it is understood that
$\rho\geq 1$.
In this context, we should mention that, adapting the results of
Ref.~\cite{HLP} based on a renormalization-group analysis in a four-generation
Majoron model, one may find a bound which is slightly more restrictive than
Eq.~(\ref{mnbound}) but still lies in the same ball park.

Furthermore, our results
depend on hadronic observables and quark-level parameters such as
the CKM-matrix elements, the quark and meson masses, the mixing angles of the
meson singlet and octet states, the pseudoscalar-meson decay constants,
and the coupling strengths of the vector mesons to the gauge bosons.
In our calculations, we use the maximum experimental values for the CKM-matrix
elements \cite{PDG} and the quark-mass values \cite{PDG,CDF}
\begin{eqnarray}
m_u&=&0.005~\mbox{GeV},\qquad m_d=0.010~\mbox{GeV},\qquad
m_s=0.199~\mbox{GeV},
\nonumber \\
m_c&=&1.35~\mbox{GeV},\qquad m_b=4.3~\mbox{GeV},\qquad m_t=176~\mbox{GeV}.
\end{eqnarray}
We keep all quark masses finite, since, {\it e.g.}, the $c$-quark and
$t$-quark contributions to the box amplitudes turn out to be comparable.
The mixing angle for vector-meson nonet states may be determined from the
quadratic Gell-Mann--Okubo mass formula to be $\theta_V=41.3^\circ$.
We treat $\theta_P$ as a free parameter because its value is not yet well
established \cite{ZielinskiRosner,thetaexp}.
For the most part, we use $\theta_P=-23^0$, the value extracted from
$e^+e^-\to e^+e^-\gamma\gamma^*\to e^+e^-(P\to\gamma\gamma)$
experiments \cite{thetaexp}. This value is consistent with a
previous analysis \cite{ZielinskiRosner}.
For the pseudoscalar-meson decay constants, we use the experimental
values \cite{PDG,thetaexp},
\begin{eqnarray}
f_{\pi^\pm}&=&92.4~\mbox{MeV},\qquad f_{K^\pm}\ =\ 113~\mbox{MeV},
\nonumber \\
f_{\pi^0}&=&84.1~\mbox{MeV},\qquad f_\eta\ =\ 94~\mbox{MeV},\qquad
f_{\eta'}\ =\ 89.1~\mbox{MeV},
\end{eqnarray}
and exploit SU(3) flavour symmetry,
\begin{equation}
f_{K^0}\ =\ f_{\bar{K}^0}\ \approx\ f_{K^\pm}.
\end{equation}
The constants $\gamma_{\tilde{V}}$ are partly extracted from the
$\tilde{V}\to e^+e^-$ decay rates, with the result that
\begin{equation}
\label{gamma}
\gamma_{\rho^0}=2.519,\qquad \gamma_\omega=2.841,\qquad
\gamma_\phi=3.037,
\end{equation}
and partly estimated assuming SU(3) symmetry: we put
$\gamma_{K^{*0}}=\gamma_{\rho^0}$ because $K^{*0}$ and $\rho^0$ are
members of the same SU(3) octet, while $\phi$ and $\omega$ are
mixtures of octet and singlet states.
Notice that all $\gamma_{\tilde{V}}$ values in Eq.~(\ref{gamma})
are very similar in size.

Having specified our input parameters, we will now discuss our numerical
results.
The widths for the decays with $K^0$, $\bar{K}^0$,
$K^{0*}$, or $\bar{K}^{0*}$ in the final state only receive contributions
from box diagrams. The branching ratios for these decays
are found to be always smaller than $10^{-14}$, that is, much
smaller than present experimental sensitivities ($\sim 10^{-6}$),
rendering these decay modes uninteresting from the experimental point of view.
Therefore, we will not pursue their study any further.

For definiteness, we will consider decays of the form
$\tau^-\to e^-M^0$ only---we set $(s_L^{\nu_\mu})^2\approx0$ to
satisfy the third inequality in Eq.~(\ref{upli}). Of course,
our estimates are also valid for the $\tau^-\to\mu^-M^0$ decays with
$(s_L^{\nu_e})^2=0$.
Our results for the branching ratios
$B(\tau^-\to e^-\pi^0/\eta/\eta'/\rho^0/\phi/\omega)$ are illustrated
in Figs.~2--6. Each figure describes the dependence of the branching
ratios on two of the free parameters, one varied
continuously and the other one in a discrete manner. All other parameters
are kept fixed.
Figure~2 shows the dependence of the branching ratios on
$m_N=m_{N_1}=m_{N_2}$ and $(s_L^{\nu_\tau})^2$.
The most promising modes are $\tau^-\to e^-\phi$, $\tau^-\to e^-\rho^0$, and
$\tau^-\to e^-\pi^0$, which, for maximum values of $m_N$ and
$(s_L^{\nu_\tau})^2$, reach branching fractions
\begin{eqnarray}
B(\tau^-\to e^-\phi)&\stackrel{\displaystyle <}{\sim}&1.6\cdot 10^{-6},
                                \nonumber\\
B(\tau^-\to e^-\rho^0)&\stackrel{\displaystyle <}{\sim}&0.9\cdot 10^{-6},
                                \nonumber\\
B(\tau^-\to e^-\pi^0)&\stackrel{\displaystyle <}{\sim}&1.0\cdot 10^{-6}.
\end{eqnarray}
This has to be compared with the present experimental bounds
\cite{CLEO,Albrecht92}
\begin{eqnarray}
B(\tau^-\to e^-\rho^0)&<&4.2\cdot 10^{-6},\nonumber\\
B(\tau^-\to e^-\pi^0)&<&1.4\cdot 10^{-4}, \nonumber\\
B(\tau^-\to \mu^-\pi^0)&<&4.4\cdot 10^{-5},
\end{eqnarray}
at the 90 \% CL.
Unfortunately, $B(\tau^-\to e^-\phi)$ has not been measured yet.
We conclude that an experimental investigation of $\tau^-\to e^-\phi$ and
a more precise determination of $B(\tau^-\to e^-\pi^0)$ and
$B(\tau^-\to \mu^-\pi^0)$ would be highly desirable.
In the high-$m_{N_1}$ limit, Fig.~2 shows the quadratic $m_N$
dependence for all branching ratios, except for $B(\tau^-\to e^-\omega)$.
In the 't~Hooft--Feynman gauge, this behaviour originates mainly from
the $Z$-boson amplitudes, $F_Z^{\tau l'}$.

At this stage, some important comments are in order.
For fixed $(s^{\nu_i}_L)^2$ values, $N_1$ and $N_2$ do not
decouple from our theory as their masses become large as compared
to $M_W$~\cite{IP,AP}. As has been mentioned in the Introduction,
$s^{\nu_i}_L \propto m_D/m_M\propto m_D/m_{N_i}$, and this nondecoupling
feature can be traced to the large SU(2)$_L$ Dirac components,
$m_D$, present in our model~\cite{AP}. Obviously, if we fix $m_D$
and take the limit $m_N\to \infty$, the heavy neutrinos will
decouple from our low-energy processes, leading to vanishing
effects~\cite{SS}. This will be illustrated in greater detail in
Figs.~4 and~5. However, for heavy neutrinos, with masses in the 1--10~TeV
range, there will be an interesting nondecoupling ``window" arising
from potentially large Dirac mass terms $m_D$. It is precisely
this nondecoupling ``window" which we are exploiting here to make
our effects sizeable.

In the case of $B(\tau^-\to e^-\pi^0/\eta/K^0)$,
we recover the expressions of Ref.~\cite{GGV} for the Dirac-neutrino
scenario if we omit the nondecoupling terms proportional to $m_N^2$ in
Eq.~(\ref{formfactors}) and in the $Z$-boson-mediated amplitudes.
The results of Ref.~\cite{GGV} are comparable to ours for $m_N$
of order $M_W$, but they fall short of our results by up to factor of 50 for
$m_N$ in the TeV region. In the case of $\omega$ production, there is a
destructive effect between logarithmic and quadratic $m_N$-dependent
nondecoupling terms coming from photon and $Z$-boson-mediated amplitudes,
respectively, and meson-mixing effects, which show up as a minimum of
the branching ratio for $m_{N_1}\approx1.6$~TeV.

We now turn to genuine Majorana-neutrino quantum effects.
Figure~3 displays the dependence of the branching fractions on the ratio
$m_{N_2}/m_{N_1}$ for the fixed values $m_{N_1}=1$~TeV and 0.5~TeV.
We emphasize that, just like in the lepton case \cite{IP},
$B(\tau^-\to e^-\pi^0/\eta/\eta'/\rho^0/\phi)$ assume their maximum values for
$m_{N_2}/m_{N_1}=2$--4 rather than in the Dirac scenario, $m_{N_1}=m_{N_2}$.
The only exception is the decay $\tau^-\to e^-\omega $, where
the maximum value is shifted to larger values of $m_{N_2}/m_{N_1}$,
of order 20, due to the accidental cancellations mentioned above.

Figures~4 and 5 illustrate the dependence of the branching ratios on
$(s_L^{\nu_\tau})^2$ and $(s_L^{\nu_e})^2$ in the heavy-Dirac-neutrino
scenario with $m_{N_1}=m_{N_2}=4$~TeV.
As may be seen in Fig.~4, the $(s_L^{\nu_\tau})^2$ dependence
of $B(\tau^-\to e^-\pi^0/\eta/\eta'/\rho^0/\phi/\omega)$ is quadratic
over the most part of the $(s_L^{\nu_\tau})^2$ range and for
any of the $(s_L^{\nu_e})^2$ values considered.
{}From Fig.~5 we see that the $(s_L^{\nu_e})^2$ dependence of
$B(\tau^-\to e^-\pi^0/\eta/\eta'/\rho^0/\phi/\omega)$
is approximately linear
for $(s_L^{\nu_e})^2<(s_L^{\nu_\tau})^2$, while it becomes quadratic for
$(s_L^{\nu_e})^2>(s_L^{\nu_\tau})^2$.
The $(s_L^{\nu_\tau})^2$ and $(s_L^{\nu_e})^2$ dependences studied above are
closely related to the decoupling behaviour of the isosinglet scale $m_M$.
As we have emphasized above, in the limit $m_N\to \infty$ for constant
$m_D$---or, equivalently, for constant $m_N$ and vanishing $m_D$, {\em i.e.},
for $(s^{\nu_i}_L)^2\to 0$---we should recover the decoupling limit, where
the branching ratios vanish as the isosinglet mass terms $m_M$ are sent to
infinity~\cite{SS}. It is then evident that the
aforementioned (non)decoupling ``window" is directly related to the
SU(2)$_L$ Dirac terms $m_D$~\cite{AP} and is reflected in
the actual $(s_L^{\nu_\tau})^2$ and $(s_L^{\nu_e})^2$ dependences seen in
Figs.~4 and 5.

In Fig.~6, we plot $B(\tau^-\to e^-\eta/\eta')$ versus $m_N=m_{N_1}=m_{N_2}$,
assuming in turn the unmixed case ($\theta_P=0$) and $\theta_P=-10^\circ$.
We see that, for $\theta_P$ decreasing, $B(\tau^-\to e^-\eta')$ increases
considerably, while $B(\tau^-\to e^-\eta)$ grows just slightly.
This illustrates that it is important to allow for nonvanishing $\theta_P$
in realistic calculations.
It is also interesting to observe that,
if tan$\theta_{\tilde{V}}=1/(\sqrt{2}c_{2W}$), the dominant nondecoupling
terms proportional to $m^2_N$ are quenched in $B(\tau^-\to l^-\omega)$.

\vspace{2cm}

\section{Conclusions}
\setcounter{equation}{0}

Motivated by the recent experimental search for lepton-number/flavour-violating
semileptonic $\tau$-lepton decays \cite{CLEO}, which are strictly prohibited
in the SM, we have explored the potential of extensions of the SM by
heavy Dirac and/or Majorana neutrinos to account for $\tau^-\to l'^-M^0$
decays, where $l'=e,\mu$ and $M^0$ is a neutral pseudoscalar or vector
meson, with branching ratios which are in line with the experimental results.
Since such models predict appreciable branching fractions for
lepton-flavour/number-violating leptonic decays of the $\tau$ lepton \cite{IP},
they are also expected to be promising candidates for explaining the problem
at hand.
In fact, we have found branching fractions in excess of $10^{-6}$ for the
channels $\tau^-\to e^-\phi$, $\tau^-\to e^-\rho^0$, and $\tau^-\to e^-\pi^0$.
Our value for $B(\tau^-\to e^-\rho)$ is comparable to the present experimental
sensitivity \cite{CLEO}.
Unfortunately, the experimental upper limit on $B(\tau^-\to e^-\pi^0)$
still exceeds our result by two orders of magnitude \cite{Albrecht92}.
For some reason, the decays $\tau^-\to e^-\phi$ or $\tau^-\to \mu^-\phi$,
which prevail in our numerical estimates, have not yet been studied
experimentally.
At this point, we should like to encourage our experimental colleagues to
undertake a search for this decay channel.

An important feature of our model is that the $\tau^-\to l'^-M^0$ decay
amplitudes exhibit a quadratic dependence on the heavy-neutrino masses,
$m_{N_1}$ and $m_{N_2}$. This nondecoupling dependence is closely related
to the large SU(2)$_L$-breaking Dirac terms $m_D$ that are allowed to
be present in our minimal {\em three}-generation seesaw-type
scenario~\cite{IP,AP}.
These $m^2_{N_i}$ terms are negligible for neutrino masses below
200~GeV, but they are dominant in the TeV region, where they may lead to
an enhancement by a factor of 50 of the respective analysis with these terms
omitted \cite{GGV}.
The same nondecoupling terms give rise to a $m_{N_2}/m_{N_1}$ dependence of
the $\tau^-\to l'^-M^0$ decay amplitudes which is
similar to the one encountered
for the decays $\tau^-\to e^+e^-e^-$, {\it etc.} \cite{IP}.
In particular, semileptonic branching ratios take their maximum values for
$m_{N_2}/m_{N_1}\approx2$--4.
The $\tau^-\to l'^-\omega$ decay rate is unobservably small
due to a destructive meson-mixing effect, which considerably
screens the dominant $Z$-exchange interaction.

The extension of the vector-meson dominance hypothesis to general vector
currents has enabled us to calculate the decays with vector mesons
in the final state.
The quark content of meson wave functions, which, for instance, is reflected
in the mixing angles, $\theta_P$ and $\theta_{\tilde{V}}$, is
also important. We have illustrated this for the production of $\eta$,
$\eta'$, and $\phi$ mesons.

\vspace{2cm}
\noindent
{\bf Acknowledgements.} We wish to thank S.\ Fajfer, D.\ Klabu\v car,
I.\ Picek, and J.\ Portoles for discussion and useful comments on
vector-meson dominance models and pseudoscalar-meson decay constants, and
A. Khodjamirian for an instructive communication concerning bilocal
hadronic matrix elements. We also thank K.\ Berkelman, R.\ Stroynowski,
and N.J.\ Urheim from CLEO for correspondence on issues related to the
decay $\tau^-\to e^-\phi$ and the interpretation of the results of
Ref.~\cite{CLEO}.
AI and AP are indebted to the Theory Group of the Max-Planck-Institut
f\"ur Physik for the kind hospitality extended to them
during a visit, when part of this work was performed.
This work is supported by the Forschungszentrum J\"ulich GmbH, Germany,
under the project number 6B0A1A.

\def\theequation{\Alph{section}.\arabic{equation}}
\begin{appendix}
\section{Appendix}

The amplitudes $a_{M^0}$, $b_{M^0}$, and $c_{M^0}$ appearing in
Eqs.~(\ref{brp}) and (\ref{brv}) may be decomposed into the form factors
$F_{Box}^{\tau l'd_ad_b}$, $F_{Box}^{\tau l'uu}$, $F_{Z}^{\tau l'}$,
$F_{\gamma}^{\tau l'}$, and $G_{\gamma}^{\tau l'}$ in the following way:
\begin{eqnarray}
a_{K^0}^{\tau l'}&=&
\frac{i\alpha^2_W}{16\, M_W^2}\,\sqrt{2}f_{K^0}\,F_{Box}^{\tau l'sd},
\nonumber\\
a_{\bar{K}^0}^{\tau l'}&=&
-\frac{i\alpha^2_W}{16\, M_W^2}\,\sqrt{2}f_{K^0}\,F_{Box}^{\tau l'ds},
\nonumber\\
a_{\pi^0}^{\tau l'}&=&
-\frac{i\alpha^2_W}{16\, M_W^2}\,
f_{\pi^0}\: [2F_{Z}^{\tau l'}+F_{Box}^{\tau l'dd}+F_{Box}^{\tau l'uu}],
\nonumber\\
a_{\eta}^{\tau l'}&=&
\frac{i\alpha^2_W}{16\, M_W^2}\,
f_\eta\: \Big[-(\frac{2c_P}{\sqrt{3}}+\frac{\sqrt{2}s_P}{\sqrt{3}})F_{Z}^{\tau
l'}
        -(\frac{c_P}{\sqrt{3}}-\frac{\sqrt{2}s_P}{\sqrt{3}})F_{Box}^{\tau l'uu}
\nonumber\\
&&      +(\frac{c_P}{\sqrt{3}}-\frac{\sqrt{2}s_P}{\sqrt{3}})F_{Box}^{\tau l'dd}
        -(\frac{2c_P}{\sqrt{3}}+\frac{\sqrt{2}s_P}{\sqrt{3}})F_{Box}^{\tau
l'ss}
       \Big],
\nonumber\\
a_{\eta'}^{\tau l'}&=&
\frac{i\alpha^2_W}{16\, M_W^2}\,
f_{\eta'}\: \Big[(\frac{\sqrt{2}c_P}{\sqrt{3}}-\frac{2s_P}{\sqrt{3}})
                  F_{Z}^{\tau l'}
        -(\frac{s_P}{\sqrt{3}}+\frac{\sqrt{2}c_P}{\sqrt{3}})F_{Box}^{\tau l'uu}
\nonumber\\
&&      +(\frac{s_P}{\sqrt{3}}+\frac{\sqrt{2}c_P}{\sqrt{3}})F_{Box}^{\tau l'dd}
        +(\frac{\sqrt{2}c_P}{\sqrt{3}}-\frac{2s_P}{\sqrt{3}})F_{Box}^{\tau
l'ss}
        \Big],
\nonumber\\
a_{K^{0*}}^{\tau l'}&=&
-\frac{i\alpha^2_W}{16\, M_W^2}\,
\frac{m_{K^{0*}}^2}{\sqrt{2}\gamma_{K^{0*}}}F_{Box}^{\tau l'sd},
\nonumber\\
a_{\bar{K}^{0*}}^{\tau l'}&=&
\frac{i\alpha^2_W}{16\, M_W^2}\,
\frac{m_{K^{0*}}^2}{\sqrt{2}\gamma_{K^{0*}}}F_{Box}^{\tau l'ds},
\nonumber\\
a_{\rho^0}^{\tau l'}&=&
\frac{i\alpha^2_W}{16\, M_W^2}\,
\frac{m_{\rho^0}^2}{\gamma_{\rho^0}}
[c_{2W}F_{Z}^{\tau l'}+\frac{1}{2}F_{Box}^{\tau l'uu}
    +\frac{1}{2}F_{Box}^{\tau l'dd}],
\nonumber\\
a_{\phi}^{\tau l'}&=&
\frac{i\alpha^2_W}{16\, M_W^2}\,
\frac{m_{\phi}^2}{\gamma_{\phi}}
\Big[(\frac{c_V}{\sqrt{3}}c_{2W}+\frac{s_V}{\sqrt{6}})F_{Z}^{\tau l'}
+(\frac{c_V}{2\sqrt{3}}-\frac{s_V}{\sqrt{6}})F_{Box}^{\tau l'uu}
\nonumber\\ &&
-(\frac{c_V}{2\sqrt{3}}-\frac{s_V}{\sqrt{6}})F_{Box}^{\tau l'dd}
+(\frac{c_V}{\sqrt{3}}+\frac{s_V}{\sqrt{6}})F_{Box}^{\tau l'ss}\Big],
\nonumber\\
a_{\omega}^{\tau l'}&=&
\frac{i\alpha^2_W}{16\, M_W^2}\,
\frac{m_{\omega}^2}{\gamma_{\omega}}
\Big[(\frac{s_V}{\sqrt{3}}c_{2W}-\frac{c_V}{\sqrt{6}})F_{Z}^{\tau l'}
+(\frac{s_V}{2\sqrt{3}}+\frac{c_V}{\sqrt{6}})F_{Box}^{\tau l'uu}
\nonumber\\ &&
-(\frac{s_V}{2\sqrt{3}}+\frac{c_V}{\sqrt{6}})F_{Box}^{\tau l'dd}
+(\frac{s_V}{\sqrt{3}}-\frac{c_V}{\sqrt{6}})F_{Box}^{\tau l'ss}\Big],
\nonumber\\
b_{K^{0*}}^{\tau l'}&=&
b_{\bar{K}^{0*}}^{\tau l'}=0,
\nonumber\\
b_{\rho^0}^{\tau l'}&=&
\frac{i\alpha^2_W s_W^2}{4\, M_W^2}\,
\frac{m_{\rho^0}^2}{2\gamma_{\rho^0}}F_{\gamma}^{\tau l'},
\nonumber\\
b_{\phi}^{\tau l'}&=&
\frac{i\alpha^2_W s_W^2}{4\, M_W^2}\,
\frac{m_{\phi}^2}{\gamma_{\phi}}\frac{c_V}{2\sqrt{3}}F_{\gamma}^{\tau l'},
\nonumber\\
b_{\omega}^{\tau l'}&=&
\frac{i\alpha^2_W s_W^2}{4\, M_W^2}\,
\frac{m_{\omega}^2}{\gamma_{\omega}}\frac{s_V}{2\sqrt{3}}F_{\gamma}^{\tau l'},
\nonumber\\
c_{K^{0*}}^{\tau l'}&=&
c_{\bar{K}^{0*}}^{\tau l'}=0,
\nonumber\\
c_{\rho^0}^{\tau l'}&=&
-\frac{i\alpha^2_W s_W^2}{4\, M_W^2}\,
\frac{m_{\rho}^2}{2\gamma_{\rho}}G_{\gamma}^{\tau l'},
\nonumber\\
c_{\phi}^{\tau l'}&=&
-\frac{i\alpha^2_W s_W^2}{4\, M_W^2}\,
\frac{m_{\phi}^2}{\gamma_{\phi}}\frac{c_V}{2\sqrt{3}}G_{\gamma}^{\tau l'},
\nonumber\\
c_{\omega}^{\tau l'}&=&
-\frac{i\alpha^2_W s_W^2}{4\, M_W^2}\,
\frac{m_{\omega}^2}{\gamma_{\omega}}\frac{s_V}{2\sqrt{3}}G_{\gamma}^{\tau l'}.
\end{eqnarray}
The form factors $F_{Box}^{\tau l'd_ad_b}$,
$F_{Box}^{\tau l'uu}$, $F_{Z}^{\tau l'}$, $F_{\gamma}^{\tau l'}$,
and $G_{\gamma}^{\tau l'}$, which also appear explicitly in Section~2,
may in turn be decomposed into elementary vertex and box functions,
$F_\gamma$, $G_\gamma$, $F_Z$, $F_{Box}$, and
$H_{Box}$.
The form factors $F_{Z}^{\tau l'}$, $F_{\gamma}^{\tau l'}$, and
$G_{\gamma}^{\tau l'}$ together with the elementary loop functions $F_\gamma$,
$G_\gamma$, $F_Z$, and $F_{Box}$ may be found in Ref.~\cite{IP}.
Here, we list $F_{Box}^{\tau l'd_ad_b}$ and $F_{Box}^{\tau l'uu}$,
\begin{eqnarray}
\label{formfactors}
F^{\tau l'uu}_{Box}&=&\sum_{i=1}^{n_R}\sum_{j=1}^{n_G}
   B_{\tau N_i}^* B_{l'N_i}V_{ud_j}^* V_{ud_j}
     \Big[H_{Box}(\lambda_{N_i},\lambda_{d_j})
   \nonumber\\ &&
     -H_{Box}(\lambda_{N_i},0)
     -H_{Box}(0,\lambda_{d_j})+H_{Box}(0,0)\Big]
   \nonumber\\ &&
  +\sum_{i=1}^{n_R}B_{\tau N_i}^* B_{l'N_i}
     [H_{Box}(\lambda_{N_i},0)-H_{Box}(0,0)],
\nonumber\\
F^{\tau l'd_a d_b}_{Box}&=&\sum_{i=1}^{n_R}\sum_{j=1}^{n_G}
    B_{\tau N_i}^* B_{l'N_i}V_{u_jd_a} V_{u_jd_b}^*
     \Big[F_{Box}(\lambda_{N_i},\lambda_{u_j})
   \nonumber\\ &&
     -F_{Box}(\lambda_{N_i},0)
     -F_{Box}(0,\lambda_{u_j})+F_{Box}(0,0)\Big]
   \nonumber\\ &&
  +\delta_{d_a d_b}\sum_{i=1}^{n_R}B_{\tau N_i}^* B_{l'N_i}
     [F_{Box}(\lambda_{N_i},0)-F_{Box}(0,0)].
\end{eqnarray}
To our knowledge, the box function $H_{Box}$ may not be found elsewhere in
the literature. After a straightforward calculation, we obtain
\begin{eqnarray}
H_{Box}(x,y)&=&\frac{1}{x-y}\Big[(4+\frac{xy}{4})
        \Big(\frac{1}{1-x}+\frac{x^2\, \ln\,x}{(1-x)^2}
            -\frac{1}{1-y}-\frac{y^2\, \ln\,y}{(1-y)^2}\Big)
   \nonumber\\ &&
        -2xy\Big(\frac{1}{1-x}+\frac{x\: \ln\,x}{(1-x)^2}
                -\frac{1}{1-y}-\frac{y\: \ln\,y}{(1-y)^2}\Big)\Big].
\end{eqnarray}
For the reader's convenience, we evaluate $H_{Box}$ for some special
arguments,
\begin{eqnarray}
H_{Box}(x,x)&=&\frac{x^3-15x^2+16x+16}{4(1-x)^2}
            +\frac{-3x^3-4x^2+16x}{2(1-x)^3}\: \ln\,x,
\nonumber\\
H_{Box}(1,x)&=&\frac{5x^2-39x+16}{8(1-x)^2}
              -\frac{2x^3+16x^2}{8(1-x)^3}\; \ln\,x,
\nonumber\\
H_{Box}(0,x)&=&\frac{4}{1-x}+\frac{4x\: \ln\,x}{(1-x)^2},
\nonumber\\
H_{Box}(0,0)&=&4,\qquad H_{Box}(0,1)=2,\qquad H_{Box}(1,1)=\frac{7}{4}.
\end{eqnarray}

\newpage

\centerline{\Large\bf Figure Captions}

\newcounter{fig}
\begin{list}{\bf\rm Fig.~\arabic{fig}: }{\usecounter{fig}
\labelwidth1.6cm \leftmargin2.5cm \labelsep0.4cm \itemsep0ex plus0.2ex }

\item Feynman graphs pertinent to the semileptonic
lepton-flavour-violating decays $\tau^-\to l'^-M^0$.

\item Branching ratios versus heavy-neutrino mass
$m_N=m_{N_1}=m_{N_2}$ for the decays $\tau^-\to e^-\pi^0$
(solid line), $\tau^-\to e^-\eta$ (dashed line), $\tau^-\to e^-\eta'$
(dotted line), $\tau^-\to e^-\rho^0$ (dot-dashed line),
$\tau^-\to e^-\phi$ (dot-long-dashed line), and
$\tau^-\to e^-\omega$ (thick-dotted line), assuming $(s^{\nu_e}_L)^2=0.01$,
$0.02\le(s^{\nu_\tau}_L)^2\le0.05$, $\theta_P=-23^0$, and $\theta_V=41.3$.

\item Branching ratios versus ratio $m_{N_2}/m_{N_1}$ for
the decays of Fig.~2, assuming $m_{N_1}=1$~TeV (0.5~TeV),
$(s^{\nu_e}_L)^2=0.01$, $(s^{\nu_\tau}_L)^2=0.05$, $\theta_P=-23^0$,
and $\theta_V=41.3$.

\item Branching ratios versus $(s^{\nu_\tau}_L)^2$ for
the decays of Fig.~2, assuming $m_{N_1}=m_{N_2}=4$~TeV,
$0.005\le(s^{\nu_e}_L)^2\le0.014$, $\theta_P=-23^0$,
and $\theta_V=41.3$.

\item Branching ratios versus $(s^{\nu_e}_L)^2$ for
the decays of Fig.~2, assuming $m_{N_1}=m_{N_2}=4$~TeV,
$0.01\le(s^{\nu_\tau}_L)^2\le0.04$, $\theta_P=-23^0$,
and $\theta_V=41.3$.

\item $B(\tau^-\to e^-\eta/\eta')$ versus $m_N=m_{N_1}=m_{N_2}$,
assuming $(s^{\nu_e}_L)^2=0.01$ and $(s^{\nu_\tau}_L)^2=0.04$:
$B(\tau^-\to e^-\eta)$ (solid line) and $B(\tau^-\to e^-\eta')$ (dashed line)
for the unmixed case $\theta_p=0$;
$B(\tau^-\to e^-\eta)$ (dotted line) and $B(\tau^-\to e^-\eta')$
(dot-dashed line) for the mixed case with $\theta_p=-10^0$.

\end{list}

\newpage

\noindent
{\bf Table~I:} Quark content of the pseudoscalar meson states $|M\rangle$
and field operators $M(x)$.

\medskip

\begin{tabular}{|l|l|l|}\hline
$|M\rangle$ & quark content of $|M\rangle$ & quark content of $M(x)$\\ \hline
$|K^+\rangle$ & $u\bar s\sim b^\dagger_ud^\dagger_s$ & $s\bar{u}$\\
$|K^0\rangle$ & $d\bar s$ & $s\bar{d}$\\
$|\pi^+\rangle$ & $-u\bar d$ & $-d\bar{u}$\\
$|\pi^0\rangle$ & $\frac{1}{\sqrt{2}}(u\bar u-d\bar d)$ &
    $\frac{1}{\sqrt{2}}(u\bar{u}-d\bar{d})$\\
$|\pi^-\rangle$ & $d\bar u$ & $u\bar{d}$\\
$|K^-\rangle$ & $s\bar u$ & $u\bar{s}$\\
$|\bar{K}^0\rangle$ & $-s\bar d$ & $-d\bar{s}$\\
$|\eta_8\rangle$ & $\frac{1}{\sqrt{6}}(u\bar u+d\bar d-2s\bar s)$ &
    $\frac{1}{\sqrt{6}}(u\bar{u}+d\bar{d}-2s\bar{s})$\\
$|\eta_1\rangle$ & $\frac{1}{\sqrt{3}}(u\bar u+d\bar d+s\bar s)$ &
    $\frac{1}{\sqrt{3}}(u\bar{u}+d\bar{d}+s\bar{s})$\\ \hline
$|\eta\rangle$ & $\cos\theta_P|\eta_8\rangle -\sin\theta_P|\eta_1\rangle$ &
    $\cos\theta_P\eta_8(x)-\sin\theta_P\eta_1(x)$\\
$|\eta'\rangle$ & $\sin\theta_P|\eta_8\rangle +\cos\theta_P|\eta_1\rangle$ &
    $\sin\theta_P\eta_8(x)+\cos\theta_P\eta_1(x)$\\ \hline
\end{tabular}

\end{appendix}
\end{document}